\documentclass[a4paper,10pt]{article}
\usepackage{authblk}% do formatowania afiliacji
\usepackage[utf8]{inputenc}
\usepackage{indentfirst}
\usepackage{graphicx}
\usepackage[english]{babel}
\usepackage{amsfonts}
\usepackage{float}
\usepackage{hyperref}
\usepackage{wrapfig}
\usepackage{caption}
\usepackage{subcaption}
\usepackage[normalem]{ulem}
\usepackage{amsmath}
\usepackage{cite}
\usepackage{enumitem}%allows to not have an indent

\usepackage{rotating}
\title{Sound emitted by some grassland animals as an indicator of motion in the surroundings}
\author[1]{M.~Pietrow\thanks{corresponding author; email: mrk@kft.umcs.lublin.pl}}
\author[2]{P.~S{\l}omski}
\author[3]{B.~Tkaczyk}
\author[4]{G.~Grzywaczewski}
\author[3]{J.~Malinowski}
\author[1]{P.~Konkol}
\affil[1]{Institute of Physics, M. Curie-Sk{\l}odowska University, ul. Pl. M. Curie-Sk{\l}odowskiej 1, 20-031 Lublin, Poland}
\affil[2]{Geographic Information Systems Development Company \emph{Martinex}, ul. Me{\l}giewska 95, 21-040 \'{S}widnik, Poland}
\affil[3]{Olympic special-interest group for young physicists in Lublin of P.~Kononowicz, ul.~B.~Pa\'snikowskiego 6, 20-707 Lublin, Poland}
\affil[4]{Department of Zoology, Animal Ecology and Wildlife Management, University of Life Sciences in Lublin, Akademicka 13, 20-950 Lublin, Poland}

\date{\today}
\begin{document}
\maketitle
\section{Abstract}
It is argued based on the results of both numerical modelling and the experiments performed on an artificial substitute of a meadow that the sound emitted by animals living in a dense surrounding such as a meadow or shrubs can be used as a tool for detection of motion. Some characteristics of the sound emitted by these animals, e.g.~its frequency, seem to be adjusted to the meadow density to optimize the effectiveness of this skill.\\
This kind of sensing the environment could be used as a useful tool improving detection of mates or predators. A study thereof would be important both from the basic-knowledge and ecological points of view (unnatural environmental changes like increasing of a noise or changes in plants species composition can make this sensing ineffective).
%%%%%%%%%%%%%%%%%%%%%%%%%%%%%%%%%%%%%%%%%%%%%%%%%%%%%%%%%%%%%%%%%%%%%%%%%%%%%%%%%%%%%%%%%%%%%%%%%
%%%%%%%%%%%%%%%%%%%%%%%%%%%%%%%%%%%%%%%%%%%%%%%%%%%%%%%%%%%%%%%%%%%%%%%%%%%%%%%%%%%%%%%%%%%%%%%%%
%%%%%%%%%%%%%%%%%%%%%%%%%%%%%%%%%%%%%%%%%%%%%%%%%%%%%%%%%%%%%%%%%%%%%%%%%%%%%%%%%%%%%%%%%%%%%%%%%
\section{Motivation}
\begin{figure}
\centering
\includegraphics[scale=0.4]{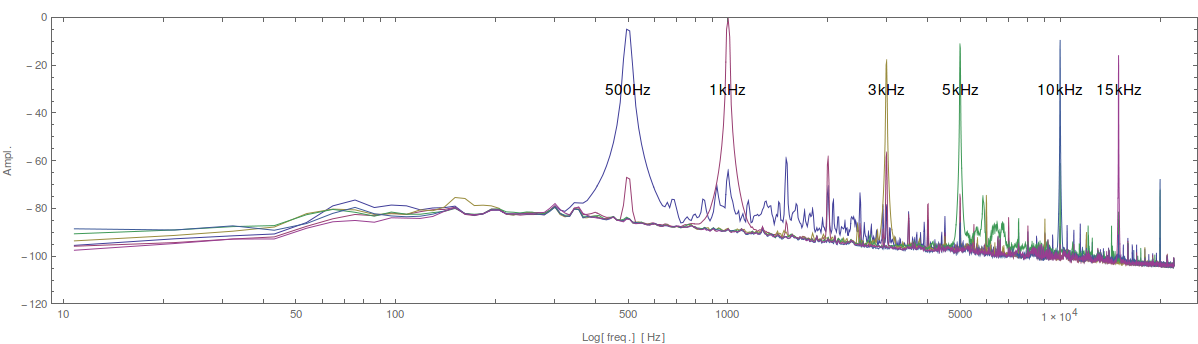}
\caption{Frequency spectrum of sound reflected from a 1$\times$1~m$^2$ wall made of cane blades exposed to one of the given tones from a range of 0.5~kHz--15~kHz (the abscissa in a logarithmic scale). The frequency of a stimulated tone is indicated by a tag near a respective peak present also in this reflected sound spectrum.}
\label{fig:Slomianka}
\end{figure}

\begin{figure}[H]
\centering
\begin{subfigure}{.5\textwidth}
  \centering
  \includegraphics[width=0.9\linewidth]{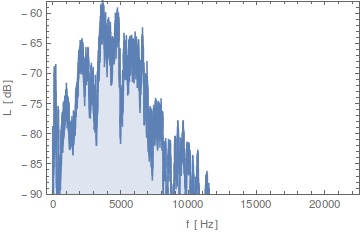}
  \caption{}
%  \label{fig:sub1}
\end{subfigure}%
\begin{subfigure}{.5\textwidth}
  \centering
  \includegraphics[width=0.9\linewidth]{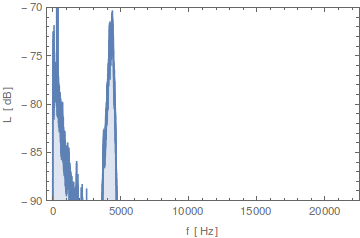}
  \caption{}
%  \label{fig:sub2}
\end{subfigure}
\caption{Sound frequency spectrum of a corncrake (a) and a cricket (b).}
\label{fig:WidmoCzestosci}
\end{figure}

A sound emitted by most animals can be heard by them~\cite{InsectHearingBook2} and could be used in many ways~\mbox{\cite{Blumstein07,Fenton84,Kroodsma91,Michelsen74,Podos04,Cocroft05}}.\\
In particular, it is interesting to consider the significance of a voice emitted by creatures living in a grassland or shrubbery. An intriguing question arises whether a long-lasting and monotonous sound emitted by some orthopterans like crickets, cicadas or some birds like a corncrake could be designed for any purpose and, if yes, whether it requires that the sound should have some special characteristics.\\
The range of vision is restricted and not effective in the grass environment. Receiving a sound feed-back from the surroundings can potentially play an important role in environment sensing.\\
Orthopterans are one of the noisiest animals living in grass and shrubs. Studies on their acoustic skills (both emission of the sound and hearing) are divided into at least into two branches: neurobiological~\mbox{\cite{Pollack00,Geurten13,Moiseff83,Huber84,Fletcher79}} and the behavioural ones.
From the latter view point, cricket's sound is identified mostly as mate attraction (\textit{phonotaxis})~\mbox{\cite{Robinson02,Poulet05,Mhatre07,Witney11}}.
It is known that crickets can recognize the direction from which the sound comes~\mbox{\cite{Wandler90}}. It also seems that some performance is preferred as a steering signal~\mbox{\cite{Meckenhaeuser13,Stabel89}}.
Crickets are able to adjust their own sound to the sound emitted by neighbours~\mbox{\cite{Walker69}}.
It is not clear yet if the sound emission by crickets has other behavioural meaning as, for example, the predator detection~\mbox{\cite{Covey12}}.
To our knowledge, the only known auditory behaviour of crickets against predators is some reaction to stimulation by ultrasounds emitted by insectivorous bats~\mbox{\cite{Doherty85,Entomology}}.
A wide review of the subject of hearing and emission of the sound by insects can be found in~\mbox{\cite{Robinson02,InsectHearingBook,Pollack17,InsectHearingBook2,Entomology}}. In all described cases, the sound of crickets is loosely related to the physical properties of the environment.\\
The second large group of animals living in grasses are birds. Although the main role of the songs of most singing birds is to attract a mate~\mbox{\cite{Rek11}}, the mostly exposed behavioural reason for sound emission in the case of a corncrake~\mbox{\cite{Tyler96}} living in grass is territorial marking~\mbox{\cite{Osiejuk13}}. Results of studies on variation of the corncrake songs show that they are connected with the degree of aggression. No reference of the sound to the properties of the environment can be found in these studies.\\
From an acoustic view point, blades of grass in a meadow can be regarded as penduli that emit sound if they are stimulated by an acoustic wave. In this way, the waves emitted from blades spread and can interfere. Fig.~\ref{fig:Slomianka} shows a sound spectrum reflected from a wall made of dry cane \textsl{Phragmites australis} blades placed one next to each other.
The blades were stimulated by one of tones from the range of 0.5~kHz--15~kHz and a reflected sound was recorded. Its spectrum is peaked at the frequency of stimulation (however, some of these peaks are broadened considerably). Additionally, in the case of stimulation with any tone, there is visible a non-specific continuous spectrum with low frequencies in the reflected sound.\\
%Diffraction is described by Huygens principle which states that every unobstructed point on a wave-front will act as a source of secondary spherical waves. The point is considered here as an oscillator which emits a wave in reply to a stimulation by an incoming wave. In our case, the 'points' scattering sound are blades of grass. The waves emitted by series of blades propagate and interfere one with another.
The role of interference of the sound is secondary in the known mechanisms of communication used by animals, e.g. echolocation, and can be considered as a disturbance. It is negligible in open space communication and in spaces scattering sound in random directions. For example, in closed spaces of caves where bats forage, the sound is reflected from distant walls which are inclined in an irregular way to each other; interference is negligible.\\
In the case of a dense environment of uniformly located blades of grass vibrating in a common plane the interference is much more important. Here, its account is amplified by a regular arrangement of reflecting objects that are sufficiently close to each other (compared to the wave length) to be able to interchange the signal with a large amplitude and thus to form nearly a collective pattern of a reflected signal.\\
In particular, interference is not to neglect for a sound propagating through a field of blades for a typical wavelength of sound emitted by animals inhabiting grass fields with a typical density. The typical distance between grass blades on meadows in middle Europe is some centimeters. In turn, the sound spectra of a corncrake \textsl{Crex Crex} and a cricket \textsl{Gryllus campestris} are shown in fig.~\ref{fig:WidmoCzestosci}.
\begin{table}[h]
\centering
\begin{tabular}{ | c | c | c | c | c | c | c |}
\hline
$f$ [kHz] & 0.5 & 1 & 3 & 5 & 10 & 15 \\
\hline
$\lambda$ [cm] &  69 & 34 & 11 & 7 & 3 & 2 \\
 \hline
\end{tabular}
\caption{Wavelength $\lambda$ as a function of frequency $f$ of a tone for several frequencies covering an acoustic spectrum. $\lambda(f)=c/f$, where $c$=343~m/s is the speed of sound in the air at 20$^{\circ}$C.}
\label{tab:DlugoscFaliOdCzestosci}
\end{table}
\noindent Although these spectra are wide to some extent, they both are concentrated at about 5~kHz. Tab.~\ref{tab:DlugoscFaliOdCzestosci} presents the wavelength for some tones as a function of their frequency. The wavelength for 5~kHz is 7~cm, which seems to be in good correspondence to the mean nearest neighbour distance for blades in a meadow. Thus, the influence of interference should not be neglected in the case of a sound emitted into blades. The composition of the interfered waves from all blades forms a final wave picture in a space. Especially, the outgoing sound can be detected at the place of the emitting source (e.g. a cricket).\\
Both in optics and in acoustics, interference fringes are analyzed mostly in a spatial domain. However, interference can be observed in a time domain, too, forming an acoustic picture of a meadow dynamically. Although a fixed position of blades could form only a static interference pattern, the moving objects (like a mate or a predator) within a field start to make this pattern a dynamical one.\\
In this paper, we present an analysis of changes in the loudness (forming interference fringes in time) of the sound perceived by an animal emitting it and hearing it at a fixed position. It is shown here that a sound with a frequency adjusted to the density of blades could be used as a sensitive tool indicating a movement nearby.
%%%%%%%%%%%%%%%%%%%%%%%%%%%%%%%%%%%%%%%%%%%%%%%%%%%%%%%%%%%%%%%%%%%%%%%%%%%%%%%%%%%%%%%%%%
%%%%%%%%%%%%%%%%%%%%%%%%%%%%%%%%%%%%%%%%%%%%%%%%%%%%%%%%%%%%%%%%%%%%%%%%%%%%%%%%%%%%%%%%%%
%%%%%%%%%%%%%%%%%%%%%%%%%%%%%%%%%%%%%%%%%%%%%%%%%%%%%%%%%%%%%%%%%%%%%%%%%%%%%%%%%%%%%%%%%%
%%%%%%%%%%%%%%%%%%%%%%%%%%%%%%%%%%%%%%%%%%%%%%%%%%%%%%%%%%%%%%%%%%%%%%%%%%%%%%%%%%%%%%%%%%
%%%%%%%%%%%%%%%%%%%%%%%%%%%%%%%%%%%%%%%%%%%%%%%%%%%%%%%%%%%%%%%%%%%%%%%%%%%%%%%%%%%%%%%%%%
%%%%%%%%%%%%%%%%%%%%%%%%%%%%%%%%%%%%%%%%%%%%%%%%%%%%%%%%%%%%%%%%%%%%%%%%%%%%%%%%%%%%%%%%%%
%%%%%%%%%%%%%%%%%%%%%%%%%%%%%%%%%%%%%%%%%%%%%%%%%%%%%%%%%%%%%%%%%%%%%%%%%%%%%%%%%%%%%%%%%%
%%%%%%%%%%%%%%%%%%%%%%%%%%%%%%%%%%%%%%%%%%%%%%%%%%%%%%%%%%%%%%%%%%%%%%%%%%%%%%%%%%%%%%%%%%%%%%%%%%%%%%%%%%%%%
%%%%%%%%%%%%%%%%%%%%%%%%%%%%%%%%%%%%%%%%%%%%%%%%%%%%%%%%%%%%%%%%%%%%%%%%%%%%%%%%%%%%%%%%%%%%%%%%%%%%%%%%%%%%%
%%%%%%%%%%%%%%%%%%%%%%%%%%%%%%%%%%%%%%%%%%%%%%%%%%%%%%%%%%%%%%%%%%%%%%%%%%%%%%%%%%%%%%%%%%%%%%%%%%%%%%%%%%%%%
%
%%%%%%%%%%%%%%%%%%%%%%%%%%%%%%%%%%%%%%%%%%%%%%%%%%%%%%%%%%%%%%%%%%%%%%%%%%%%%%%%%%%%%%%%%%%%%%%%%%%%%%%%%%%%%
%%%%%%%%%%%%%%%%%%%%%%%%%%%%%%%%%%%%%%%%%%%%%%%%%%%%%%%%%%%%%%%%%%%%%%%%%%%%%%%%%%%%%%%%%%%%%%%%%%%%%%%%%%%%%
%%%%%%%%%%%%%%%%%%%%%%%%%%%%%%%%%%%%%%%%%%%%%%%%%%%%%%%%%%%%%%%%%%%%%%%%%%%%%%%%%%%%%%%%%%%%%%%%%\paragraph{Numerical experiments:}
%%%%%%%%%%%%%%%%%%%%%%%%%%%%%%%%%%%%%%%%%%%%%%%%%%%%%%%%%%%%%%%%%%%%%%%%%%%%%%%%%%%%%%%%%%%%%%%%%%%%%%%%%%%%%%%%%%%%%%%%%%
%%%%%%%%%%%%%%%%%%%%%%%%%%%%%%%%%%%%%%%%%%%%%%%%%%%%%%%%%%%%%%%%%%%%%%%%%%%%%%%%%%%%%%%%%%%%%%%
\section{Acoustic experiments}
\begin{figure}
\centering
\begin{subfigure}[b]{0.355\textwidth}
\includegraphics[width=\textwidth]{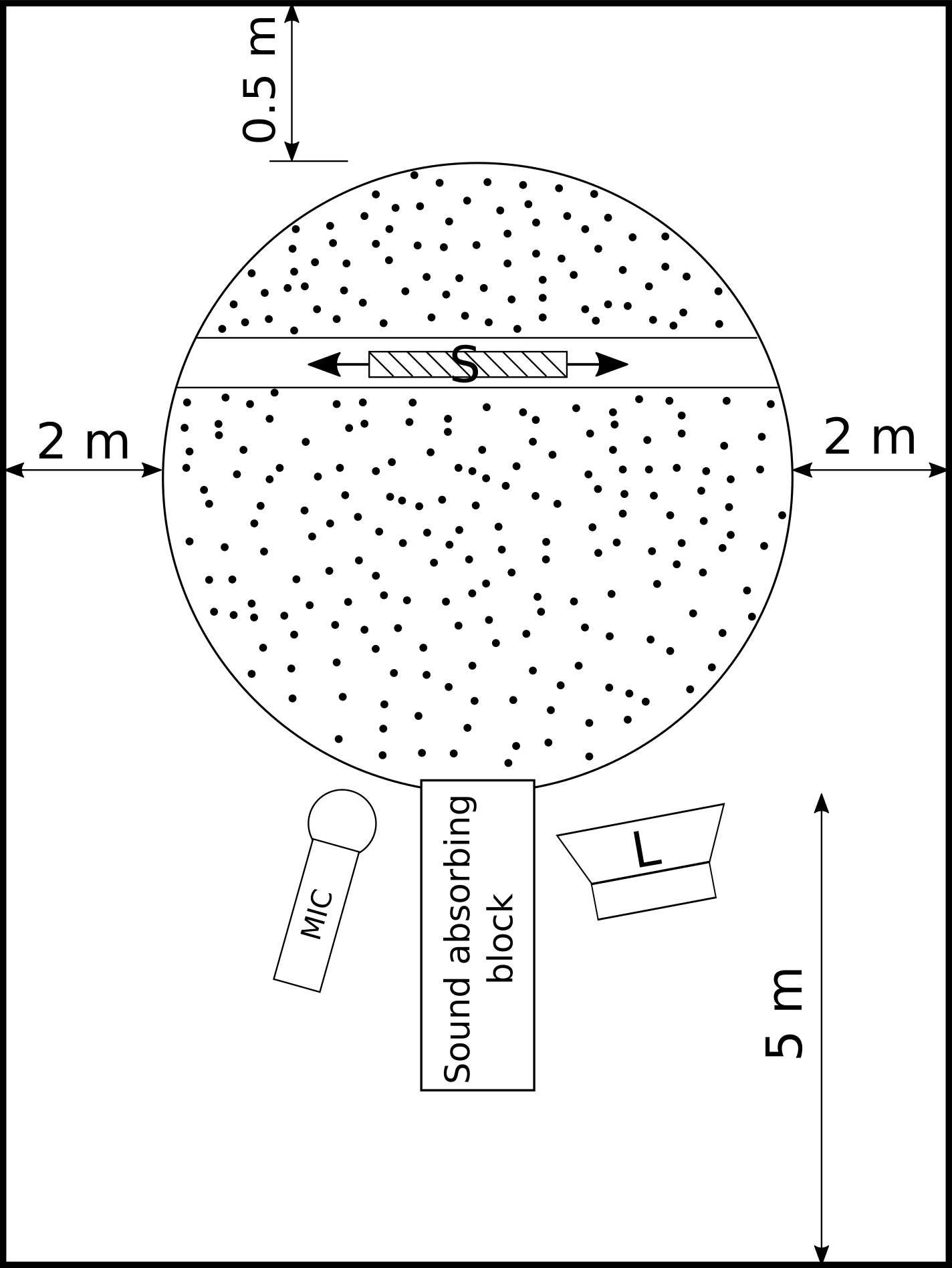}
\caption{}
\label{fig:RecordingScheme}
\end{subfigure}
\begin{subfigure}[b]{0.49\textwidth}
\includegraphics[width=\textwidth]{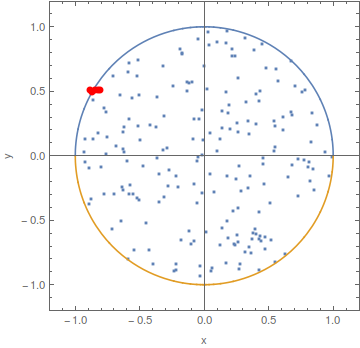}
\caption{}
\label{fig:ArtificialMeadow}
\end{subfigure}
\caption{(a)~--~Scheme of the experimental setup. '$S$' denotes a moving shutter whereas 'MIC'~--~the position of the microphone to measure the loudness of the reflected sound from the field almost in the place of the singer represented by a loudspeaker '$L$' here. (b)~--~An artificial meadow with a radius $R$=1~m made of $N$=200 randomly placed blades. Red points denote an initial newcomer position. The source emitting the sound is placed at (0,0).}
\end{figure}

\begin{figure*}[htp]
    \centering
    \begin{subfigure}{0.33\textwidth}
      \centering
      \includegraphics[width=5.5cm]{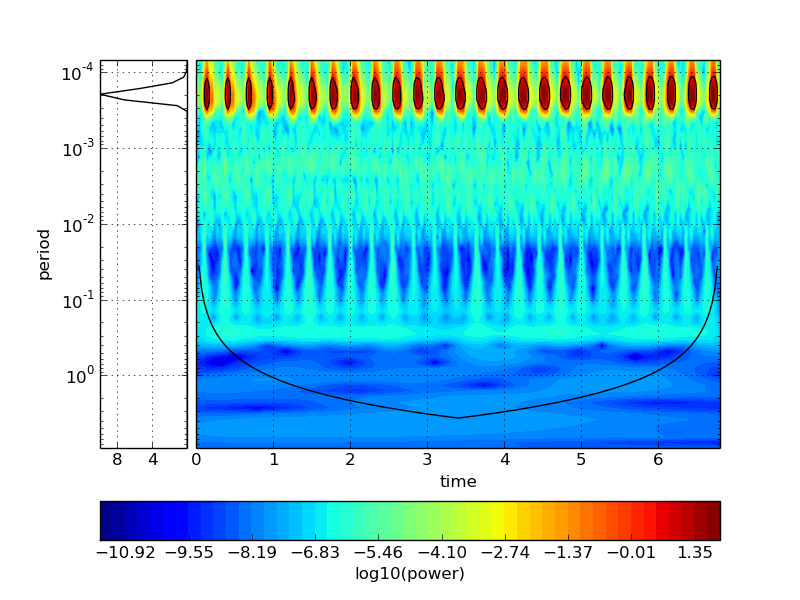}
      \caption{original sound (no blades)}
      \label{fig:MorletWavelet_Original}
    \end{subfigure}%
    \begin{subfigure}{0.33\textwidth}
      \centering
      \includegraphics[width=5.5cm]{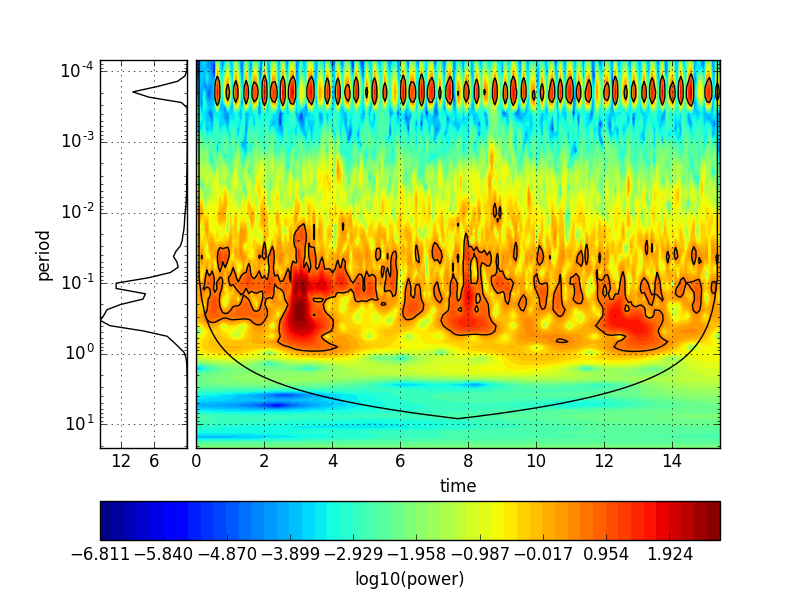}
      \caption{$\alpha$=0.06~m}
      \label{fig:MorletWavelet_n200Swierszcz}
    \end{subfigure}
    \begin{subfigure}{0.33\textwidth}
      \centering
      \includegraphics[width=5.5cm]{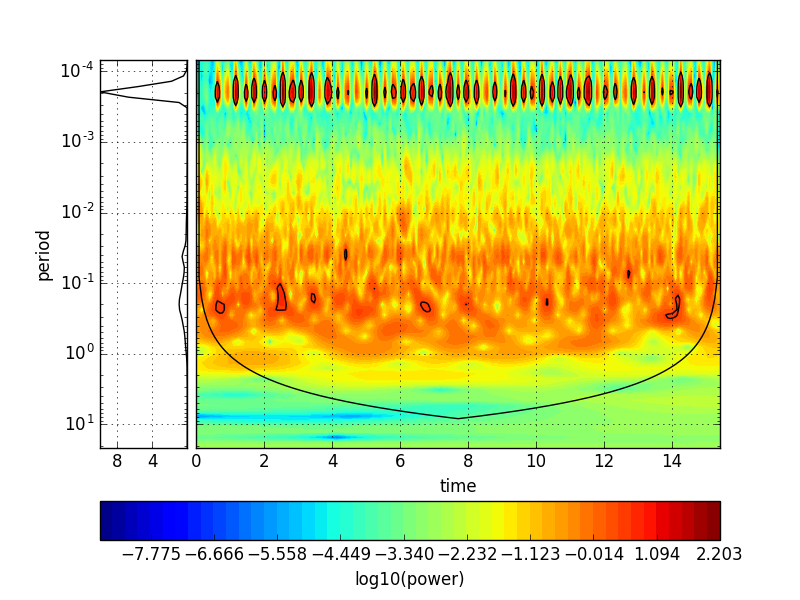}
      \caption{$\alpha$=0.04~m}
     \label{fig:MorleWavelet_n500Swierszcz}
    \end{subfigure}
  \caption {Morlet wavelet decomposition of the prepared cricket's sound reflected from the artificial meadow. At each time (on the abscissa), the oscillation period spectrum of the sound is shown. Their intensity (represented by power) is denoted by colorus. An additional chart on the left side of each plot shows the mean relative amplitude as a function of the period of the oscillations. The time series is registered while the newcomer is passing across the artificial meadow with a velocity of 0.1~m/s.}
\label{fig:MorletWavelet}
\end{figure*}

\begin{figure}
\centering
\begin{subfigure}[b]{\textwidth}
\includegraphics[width=\textwidth]{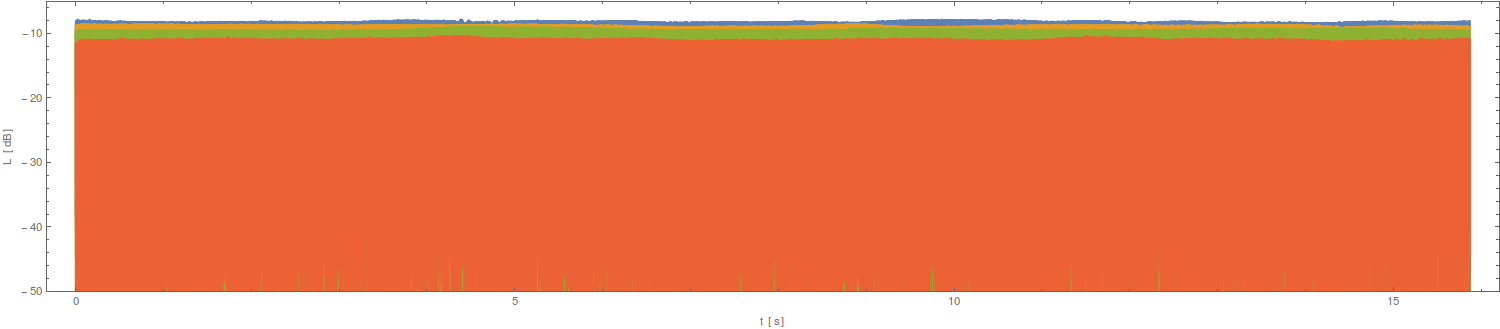}
\caption{}
\label{fig:v15_05kHz}
\end{subfigure}
\begin{subfigure}[b]{\textwidth}
\includegraphics[width=\textwidth]{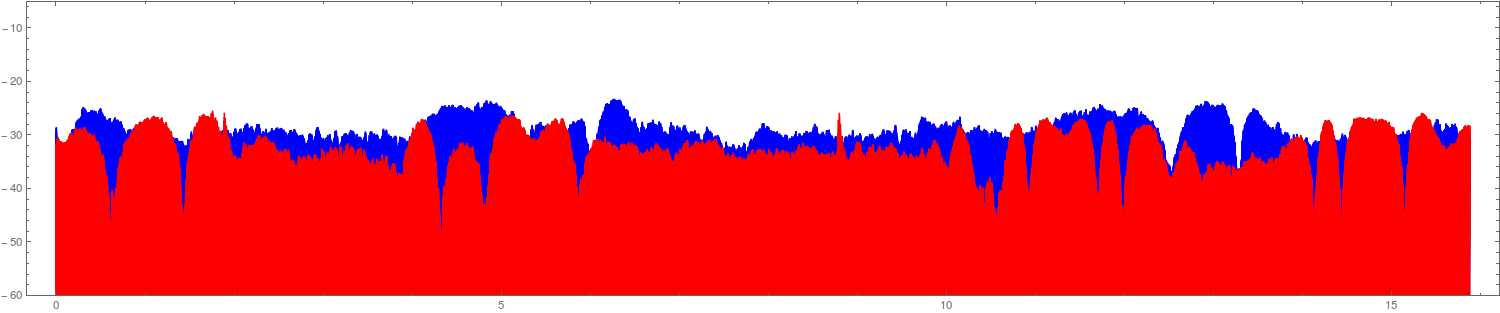}
\caption{}
\label{fig:v15_3kHz}
\end{subfigure}
\begin{subfigure}[b]{\textwidth}
\includegraphics[width=\textwidth]{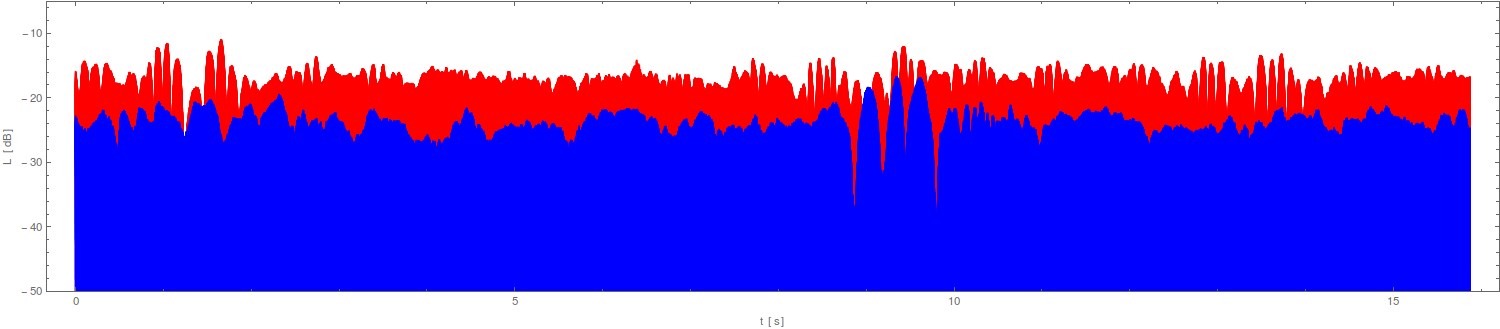}
\caption{}
\label{fig:v15_10kHz}
\end{subfigure}
\caption{Relative changes in loudness in time of the reflected sound for the artificial meadow penetrated by pure tones: (a)~--~0.5~kHz, (b)~--~3~kHz, and (c)~--~10~kHz. The velocity of the newcomer was $w$=0.1~m/s. In sub-figure (a), the colors denote various numbers of blades ($N$=0, 100, ..., 500). In (b): $\alpha$=0.3~m for the blue series, and $\alpha$=0.04~m for the red series. In (c): blue~--~$\alpha$=0.1~m, red~--~$\alpha$=0.04~m (here, to show differences more accurately, the spectrum in blue was shifted up by 5~dB).}\label{fig:v15}
\end{figure}

\begin{figure}
    \centering
    \begin{subfigure}[b]{\textwidth}
        \includegraphics[width=\textwidth]{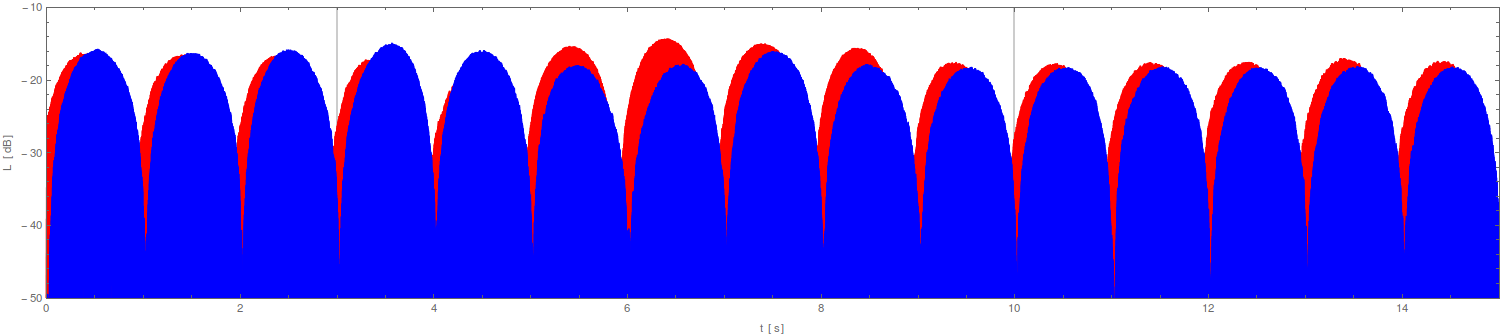}
        \caption{}
        \label{fig:05Hz1kHz}
    \end{subfigure}
    ~ %add desired spacing between images, e. g. ~, \quad, \qquad, \hfill etc. 
    \begin{subfigure}[b]{\textwidth}
        \includegraphics[width=\textwidth]{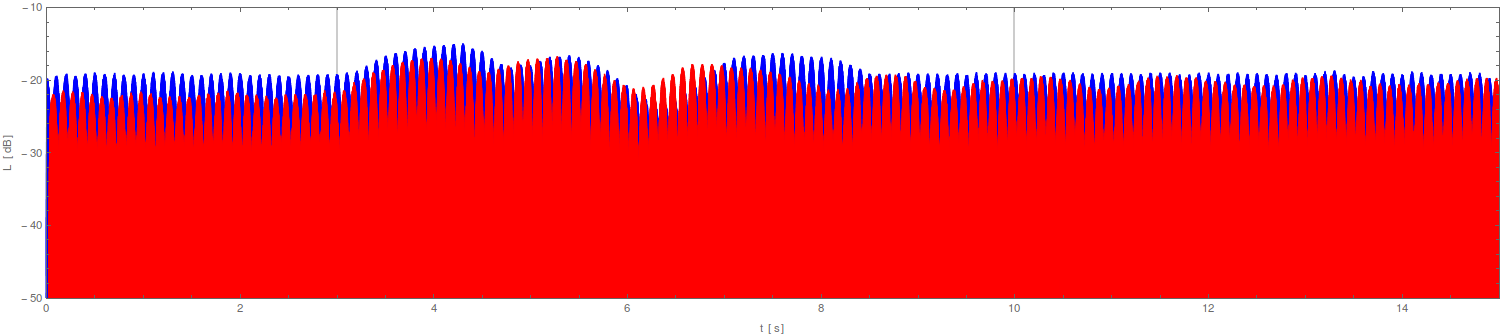}
        \caption{}
        \label{fig:5Hz1kHz}
    \end{subfigure}
\caption{A time-line of loudness for a low carrier frequency tone $f$=1~kHz scattered from the artificial meadow. The modulation frequencies are: (a) $f_m$=0.5~Hz and (b) $f_m$=5~Hz. In part (a), $\alpha$ is 0.1~m (blue) and 0.04~m (red). In (b), $\alpha$ is 0.1~m (blue) and 0.06~m (red). The gray vertical lines denote the moment of starting and stopping moving by the newcomer ($w$=0.1~m/s). In the case of the greater $f_m$, (b), more intensive fluctuations are visible during the passage of the newcomer.}
\label{fig:Mod1kHz}
\end{figure}

\begin{figure}
\centering
\includegraphics[scale=0.3]{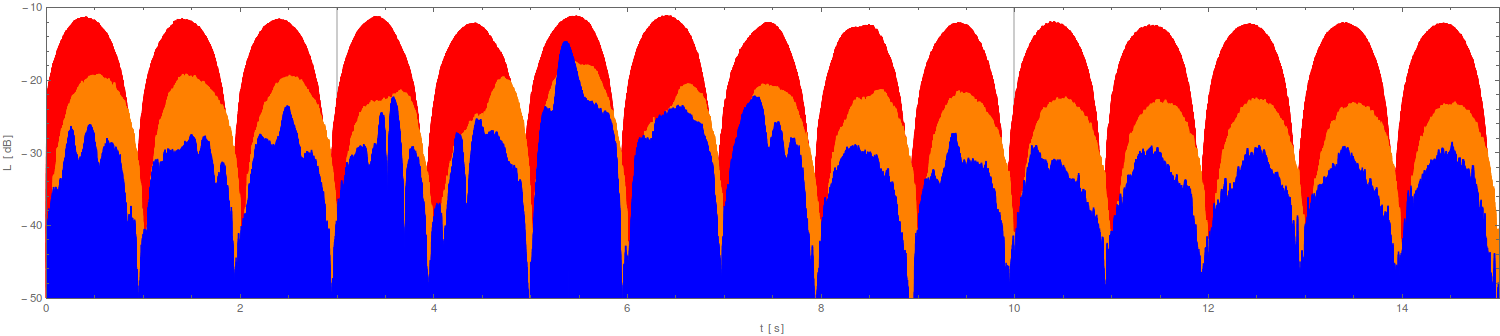}
\caption{Low modulation frequency case: a time-line of loudness of a modulated tone $f_m$=0.5~Hz $f$=5~kHz scattered from the meadow with $\alpha$=0.1~m (blue), $\alpha$=0.06~m (orange), and $\alpha$=0.04 (red). A meaning of gray vertical lines is the same as in~fig.~\ref{fig:Mod1kHz}.}
\label{fig:05Hz5kHz}
\end{figure}

\begin{figure}
\centering
\includegraphics[scale=0.3]{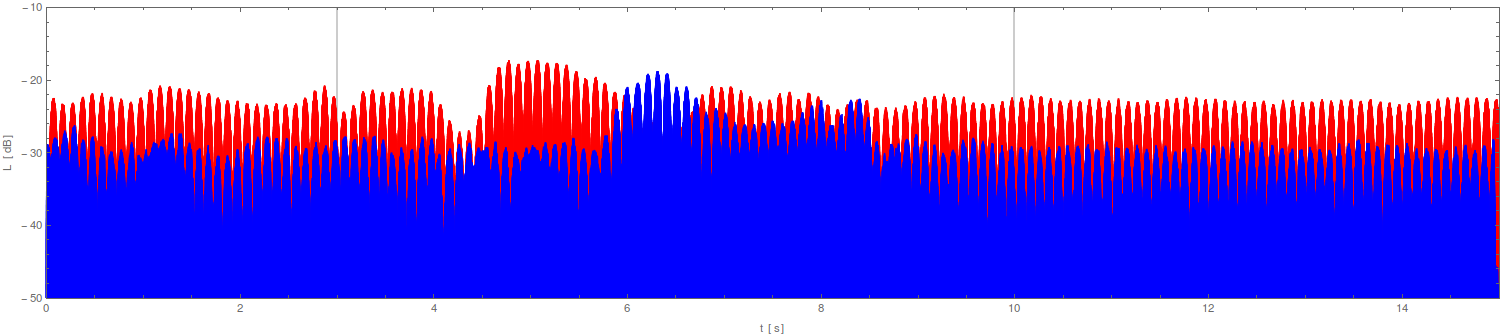}
\caption{High modulation frequency case: a time-line of a reflected signal loudness of a modulated tone $f_m$=5~Hz $f$=5~kHz from the meadow with $\alpha$=0.1~m (blue) and with $\alpha$=0.06~m (red). The red part was shifted by 5~dB to see the differences more accurately. The meaning of the gray vertical lines is the same as in~fig.~\ref{fig:Mod1kHz}. Considerable fluctuations of the sound loudness are visible when the newcomer is moving through the area with blades.}
\label{fig:5Hz5kHz}
\end{figure}

\begin{figure}
\centering
\includegraphics[scale=0.3]{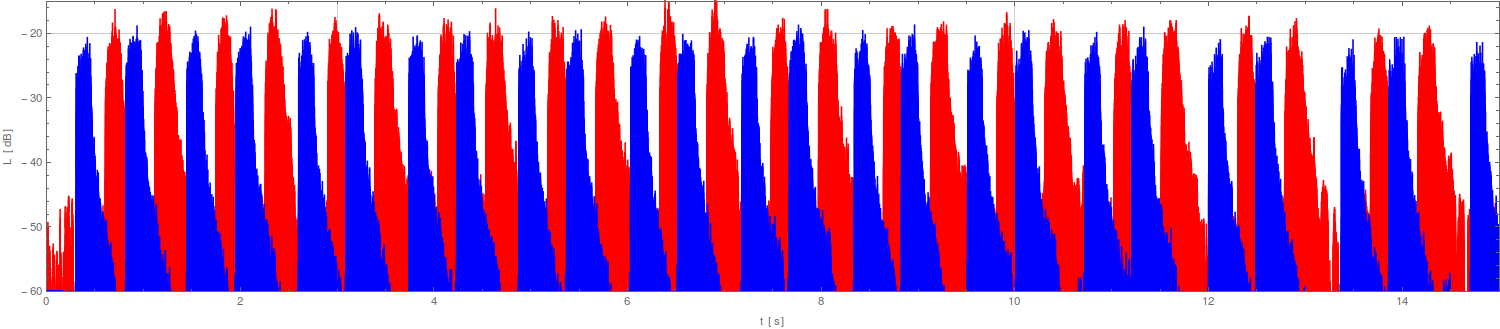}
\caption{Loudness of a reflected signal with time for the voice of a corncrake emitted from the loudspeaker; blue points~--~the original recording of the voice, red points~--~the signal emitted into the field with $\alpha$=0.06~m. Some fluctuations of the amplitude of peaks are visible during the passage of the newcomer.}
\label{fig:derkacz_OriginalVSn200}
\end{figure}

\begin{figure}
\centering
\includegraphics[width=\textwidth]{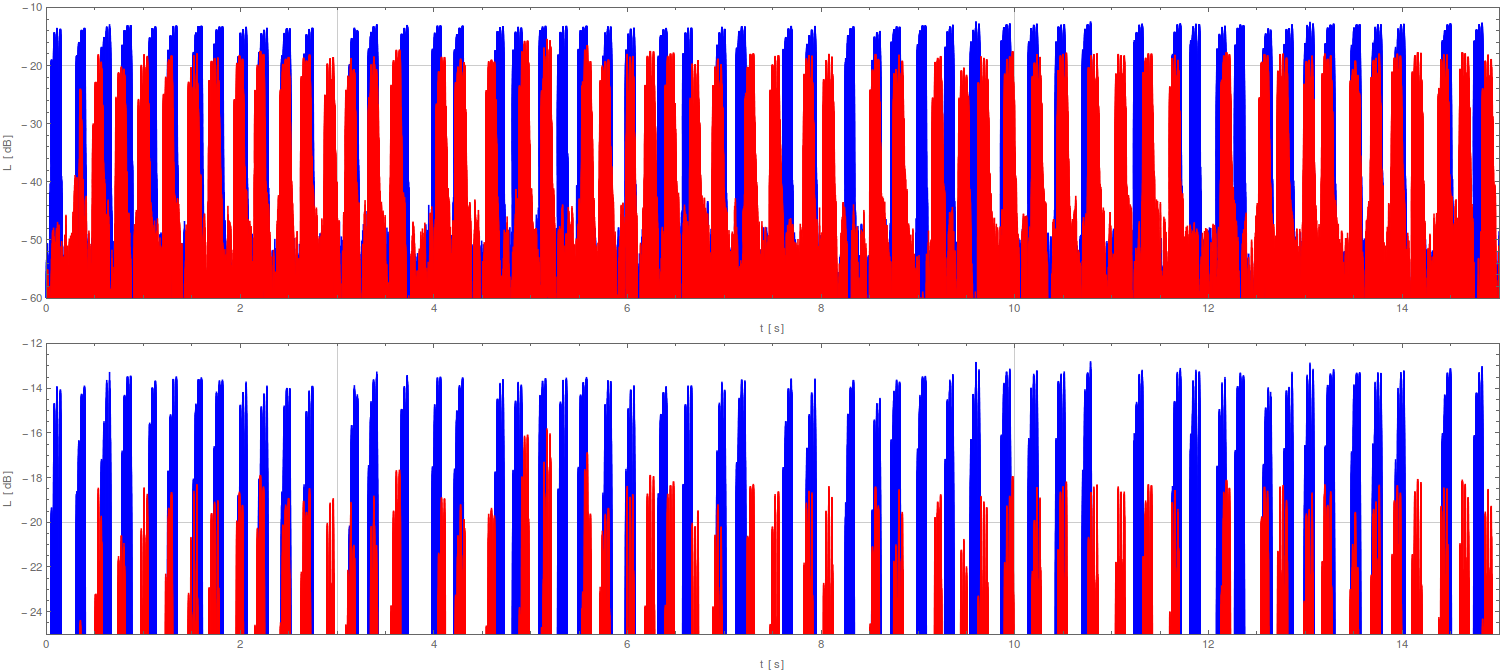}
\caption{Loudness of a reflected signal with time of the cricket's voice reflected from a meadow with $\alpha$=0.06~m (red). The blue points~--~an original signal. The lower part of the figure is a magnification of the region of the peaks. Fluctuations of the amplitude of peaks are visible.}
\label{fig:swierszcz_OriginalVSn200}
\end{figure}

%%%%%%%%%%%%%%%%%%%%%%%%%%%%%%%%%%%%%%%%%%%%%%%%%%%%%%%%%%%%%%%%%%%%%%%%%%%%%%%%%%%%%%%%%%%%%%%
%/home/marek/Dokumenty/WhatILike/WoodsAndGrasses/Numerics/c/ObliczeniaZUczniami/wavelet
%/home/marek/Dokumenty/WhatILike/WoodsAndGrasses/Numerics/c/ObliczeniaZUczniami/Dzwieki/rCentrum3/Nagrania/Swierszcz
%Besides of the Fourier analysis, the method which gives an accurate analysis of changes a sound spectrum in time is a wavelet decomposition.
%
%\paragraph{Definitions.}
\paragraph{Description of the experiments.}
The experiment analysed here consisted in emission of sound toward an artificial meadow and registering the sound returning to the place of emission thereof. To measure changes in the loudness of the sound reflected from a meadow, we constructed an experimental setup shown in fig.~\ref{fig:RecordingScheme}.\\
Hereafter, we shall use the notion \emph{artificial meadow} for a circular surface with several blades scattered randomly throughout this surface (fig.~\ref{fig:RecordingScheme}). In the case of our acoustic experiments, these blades were dry blades of cane stuck into a Styrofoam board.
\\
Furthermore, a \emph{newcomer} denotes an animal (a mate or a predator) which can move through the meadow. Its move is supposed to be detected by an animal emitting the sound by registration of loudness fluctuations of the sound reflected from the meadow. In the acoustic experiments, the newcomer was substituted by a 20$\times$30~cm metallic or Plexiglas plate moving 15~cm above the ground level between blades~--~denoted as $S$ in the fig.~\ref{fig:RecordingScheme}.\\
To calculate the number of blades needed to construct a field with a given nearest neighbour distance $\alpha$, we used formula
\begin{equation}
\alpha=\frac{R\sqrt{\pi}}{2\sqrt{N}},
\label{eq:Torquato}
\end{equation}
where $R$~--~the radius of the meadow, and $N$~--~the number of blades~\cite{Torquato02}. It is clear from eq.~(\ref{eq:Torquato}) that the surface density is an unambiguous function of $\alpha$.
%%%%%%%%%%%%%%%%%%%%%%%%%%%%%%%%%%%%%%%%%%%%%%%
%%%%%%%%%%%%%%%%%%%%%%%%%%%%%%%%%%%%%%%%%%%%%%%
%%%%%%%%%%%%%%%%%%%%%%%%%%%%%%%%%%%%%%%%%%%%%%%
%%%%%%%%%%%%%%%%%%%%%%%%%%%%%%%%%%%%%%%%%%%%%%%%%%%%%%%%%%%%%%%%%%%%%%%%%%%%%%%%%%%%%%%%%%\paragraph{\uwave{Acoustic experiments:}}
%Some preliminary experiments were performed~--~see Appendix~(sec.~\ref{sec:append1}).
%%%%%%%%%%%%%%%%%%%%%%%%%%%%%%%%%%%%%%%%%%%%%%%%%%%%%%%%%%%%%%%%%%%%%%%%%%%%%%%%%%%%%%%%%%%%%%%%%%%%%%%%%%%%%%%%%%%
%%%%%%%%%%%%%%%%%%%%%%%%%%%%%%%%%%%%%%%%%%%%%%%%%%%%%%%%%%%%%%%%%%%%%%%%%%%%%%%%%%%%%%%%%%%%%%%%%%%%%%%%%%%%%%%%%%%
%
\\
All reliable recordings that we performed were taken to a recording room of a local radio station. The walls in this room were padded with mats absorbing some fraction of sound. In our further experiments, we measured the propagation of the following sounds through the meadow:
\begin{itemize}[leftmargin=*]
\item a set of pure tones $\mathcal{F}$=\{0.5~kHz, 1~kHz, 3~kHz, 5~kHz, 10~kHz, 15~kHz\},
\item a set of modulated tones $\Upsilon=\mathcal{F}_m \times \mathcal{F}$=\{0.5~Hz, 1~Hz, 2~Hz, 5~Hz\} $\times$ \{0.5~kHz, 1~kHz, 3~kHz, 5~kHz, 10~kHz, 15~kHz\}, where $\mathcal{F}$ is a chosen set of carrier frequencies $f$, whereas $\mathcal{F}_m$ is a set of modulation frequencies $f_m$,
%\item a white noise
\item prepared samples of corncrake's and cricket's sounds.
\end{itemize}
The samples of tones and modulated tones were generated in \textit{Mathematica} software. They lasted 15~s and were sampled with a frequency of 44100~Hz (CD-quality).\\
The prepared samples of natural sounds of a corncrake and a cricket were transformed in the \emph{Audacity} software, where a given sound was cropped to one period of a chirp, and next this part was multiplied. This preparation allowed avoiding some natural irregularities present during a few-minute call.\\
The samples were emitted by a \emph{JBL EON10 G2} loudspeaker into the artificial meadow. The reflected sound was recorded with an \emph{Audio-technica AT2031} microphone coupled with a computer and with a \emph{Sennheiser ME67} microphone coupled with a \emph{Zoom H4n} handy recorder. The recorded sound was analysed by the \emph{Audacity} software and programs written in \emph{Mathematica}, \emph{C++}, and \emph{Python}.
\paragraph{Results of the experiments.}
A spectral decomposition of a received sound reveals interesting changes in time. We used wavelet analysis as a comprehensive tool to show the frequency spectrum in time. Fig.~\ref{fig:MorletWavelet} shows a Morlet wavelet~\cite{Morlet} decomposition of the prepared signal of the cricket.
%The first three seconds of the recording it is a time when an artificial newcomer stays at one of the edges of a meadow.
The sub-figures show the intensity of the signal (in colours; represented by power) at a given time (on the abscissa) and for a given frequency for three values of the mean nearest distance $\alpha$ (the period is shown instead of the frequency on the ordinate of the plot). During these recordings, the newcomer travelled across the artificial meadow with a velocity of 0.15~m/s. A comparison shows that if blades are present, there appear additional frequencies in a signal (a period 10$^0$ -- 10$^{-3}$~s) whose amplitude is less homogeneous in time for a denser meadow. The amplitude of these frequencies is the largest for an intermediate density of the meadow (fig.~\ref{fig:MorletWavelet_n200Swierszcz}). The appearance of this frequency band is characteristic for the sound of a cricket and a corncrake as well as modulated tones. Intensive fluctuations of these frequencies with time are visible.
\\
The origin of these additional frequencies is acoustic beats, which are a function of the modulation frequency, the density of blades, and the velocity of the newcomer. They can possibly be used by animals for identification of changes in the position of objects. Furthermore, some of these additional frequencies come from the effect of generation of some spectrum of frequencies by blades when they are stimulated by sound (as demonstrated in fig.~\ref{fig:Slomianka}).
%%%%%%%%%%%%%%%%%%%%%%%%%%%%%%%%%%%%%%%%%%%%%%%%%%%%%%%%%%%%%%%%%%%%%%%%%%%%%%%%%%%%%%%%%%%%%%%
%%%%%%%%%%%%%%%%%%%%%%%%%%%%%%%%%%%%%%%%%%%%%%%%%%%%%%%%%%%%%%%%%%%%%%%%%%%%%%%%%%%%%%%%%%%%%%%
% Tones
%%%v=0
%%%\begin{figure}[H]
%%%\centering
%%%\begin{subfigure}{.33\textwidth}
%%%  \centering
%%%  \includegraphics[width=0.95\linewidth]{figs/Wyniki_rCentrum2_v0_05kHz.png}
%%%  \caption{0.5 kHz}
%%%%  \label{fig:sub1}
%%%\end{subfigure}%
%%%\begin{subfigure}{.33\textwidth}
%%%  \centering
%%%  \includegraphics[width=0.95\linewidth]{figs/Wyniki_rCentrum2_v0_5kHz.png}
%%%  \caption{5 kHz}
%%%%  \label{fig:sub2}
%%%\end{subfigure}
%%%\begin{subfigure}{.33\textwidth}
%%%  \centering
%%%  \includegraphics[width=0.95\linewidth]{figs/Wyniki_rCentrum2_v0_15kHz.png}
%%%  \caption{15 kHz}
%%%%  \label{fig:sub3}
%%%\end{subfigure}
%%%\caption{Tones, $v$=0.}
%%%\label{fig:Tones_v0}
%%%\end{figure}
% /home/marek/Dokumenty/WhatILike/WoodsAndGrasses/Numerics/c/ObliczeniaZUczniami/Dzwieki/RCentrum2
%
\\
%Tones
Furthermore, we performed some measurements of the loudness changes in time for the reflected sound for some simple tones and the velocity of the newcomer $w$=0.1~m/s. For low frequency $f$=0.5~kHz, the loudness changes by 1-2~dB depended on the density (fig.~\ref{fig:v15_05kHz}). In this case, much of the intensity is reflected from the wall of the meadow and no interference effects are visible.
%%%\begin{figure}
%%%\centering
%%%\includegraphics[scale=0.3]{figs/Wyniki_rCentrum2_v15_05kHz.png}
%%%\caption{v15, 0.5~kHz, n=0,...,500.}
%%%\label{fig:v15_05kHz}
%%%\end{figure}
%%%For higher frequencies, the sound can penetrate the inside of the meadow and can come back. For 3~kHz tone (the wavelength is 11~cm), the meadow with $n$=100 -- 500 blades is sufficient to produce interference pattern (the mean nearest distance is 9-4~cm, respectively). A comparison of the signals for some cases is given in fig.~\ref{fig:v15_3kHz}
%%%\begin{figure}
%%%\centering
%%%\includegraphics[scale=0.3]{figs/Wyniki_rCentrum2_v15_3kHz_n0VSn500.png}
%%%\caption{v15 (a newcomer moves), 3~kHz, $n$=0 (blue), $n$=500 (orange).}
%%%\label{fig:v15_3kHz}
%%%\end{figure}
%%%and in fig.~\ref{fig:v15_10kHz}. The latest figure shows a result for 10~kHz emitted toward a field with $n$=0, 200 and 500 blades.
%%%\begin{figure}
%%%\centering
%%%\includegraphics[scale=0.3]{figs/Wyniki_rCentrum2_v15_10kHz_n0VSn500.png}
%%%\caption{v15 (a newcomer moves), 10~kHz ($\lambda$=3~cm), $n$=0 (orange), $n$=500 (blue). To show differences more accurately, the spectrum for $n$=500 was shifted up by 5~dB.}
%%%\label{fig:v15_10kHz}
%%%\end{figure}
%%%%%%%%%%%%%%%%%%%%%%%%%%%%%%%%%%%%%%%%%%%%%%%%%%%%%%%%%%%%%%%%%%%%%%%%%%%%%%%%%%%%%%%%%%%%%%%%%%%%%%
For greater frequencies, the sound can penetrate the inside of the meadow and it is partially dispersed. However, in this case, interference inside is possible. For example, for a tone 3~kHz (the wavelength is $\lambda$=11~cm), the condition $\alpha\simeq \lambda$ is true. It suffices to produce an interference pattern~--~fig.~\ref{fig:v15_3kHz}. A similar effect is observed for greater frequencies~--~fig.~\ref{fig:v15_10kHz}.\\
%
%%%%%%%%%%%%%%%%%%%%%%%%%%%%%%%%%%%%%%%%%%%%%%%%%%%%%%%%%%%%%%%%%%%%%%%%%%%%%%%%%%%%%%%%%%%%%%%%%%%%
%%%%%%%%%%%%%%%%%%%%%%%%%%%%%%%%%%%%%%%%%%%%%%%%%%%%%%%%%%%%%%%%%%%%%%%%%%%%%%%%%%%%%%%%%%%%%%%%%%%%
%%%%%%%%%%%%%%%%%%%%%%%%%%%%%%%%%%%%%%%%%%%%%%%%%%%%%%%%%%%%%%%%%%%%%%%%%%%%%%%%%%%%%%%%%%%%%%%%%%%%
% Modulated tones
Similar experiments were repeated for modulated tones $\Upsilon$. For these tones, we performed recordings where the newcomer (a shutter) was at rest during the first 3~s; next, it moved with a velocity of $w$=0.1~m/s from the 3$^{\text{rd}}$ to the 10$^{\text{th}}$ second, and then was at rest again from the 10$^{\text{th}}$ to the 15$^{\text{th}}$ second.\\
The result for both $f$ and $f_m$, which have relatively small values, is shown in fig.~\ref{fig:05Hz1kHz}.
For the meadow with sparse blades (a blue line), some waving of loudness is observed but it is inadequate to the moments when the shutter appears in front of the microphone. For a greater density (in red there), a more adequate increase in the amplitude is visible. Some shift of the amplitude in time, compared to the signal in blue, is an interference effect.
\\
The waving of the amplitude is present for both meadow densities if greater modulating frequency $f_m$ is applied~--~fig.~\ref{fig:5Hz1kHz}. For the dense meadow (red), the waving reduces the amplitude very slowly~--~some waving is visible even after moving out of a meadow by the shutter. The waving is the greatest when the newcomer passes nearby but it is impossible to reconstruct its position from this pattern.\\
%%%%%%%%%%%%%%%%%%%%%%%%%%%%%%%%%%%%%%%%%%%%%%%%%%%%%%%%%%%%%%%%%%%%%%%%%%%%%%%%%%%%%%%%%%%%%%%%%%%%%%%%%%
It is interesting to check the variation of loudness in time for a carrier frequency $f$ comparable to those which are characteristic for corncrakes or crickets. In fig.~\ref{fig:05Hz5kHz}, the signal is compared for three densities of the meadow. This signal seems to rebuild better with the increasing density of the meadow. The more regular the signal is the less information it carries about the passage of the newcomer. However, in the case of small density (blue), the amplitude fluctuation is significant and adequate to the position of the newcomer. These fluctuations are not as great as those giving separate chirps so they are possibly not much informative. In this case, the small density is inadequate for realistic meadows so these fluctuations possibly do not correspond to realistic case. For these meadows, a signal in red is rather expected which allows a conclusion that such a small value of the $f_m$ as in this case would be inadequate for signaling in realistic meadows. Thus, the small $f_m$ values are not appropriate to give information about the movement in the case of the dense meadow.
\\
In contrast, when the modulation frequency $f_m$ is greater, the interference picture starts to be more informative~--~fig.~\ref{fig:5Hz5kHz}. The maximum value of fluctuations in this case precedes the moment of passing the newcomer at the nearest distance to the cricket.
Furthermore, a fluctuation of the amplitude of the signal is greater in the case of greater meadow density (red). For more quantitative comparison of these signals fluctuations a \textit{distortion function} was introduced~--~see the Appendix.%~--~fig.~\ref{fig:5Hz5kHz_n0VSn200_magnified}.
%\uwave{The fluctuations are present in a case of a sparse and a dense meadow. It suggests that the fluctuation amplitude depend on a velocity of a newcomer.} The better adjustment to the velocity of the newcomer (to a distance which it passes in a scale of seconds) \uwave{the greater are fluctuations}.
%
%\begin{figure}
%\centering
%\includegraphics[scale=0.3]{figs/5Hz5kHz_rCentrum11102016_n0VSn200_part.png}
%\caption{Focus on a part of the time spectrum of the sound for $\alpha$=0.1~m (orange) and $\alpha$=0.06 (blue) for $f_m$=5~Hz $f$=5~kHz. Much greater fluctuations are visible in the case when a meadow is dense. To make the differences more visible, the blue spectrum was shifted up by 15~dB.}
%\label{fig:5Hz5kHz_n0VSn200_magnified}
%\end{figure}
%%%%%%%%%%%%%%%%%%%%%%%%%%%%%%%%%%%%%%%%%%%%%%%%%%%%%%%%%%%%%%%%%%%%%%%%%%%%%%%%%%%%%%%%%%%%%%%%%%%%%
%
\\
%%%%%%%%%%%%%%%%%%%%%%%%%%%%%%%%%%%%%%%%%%%%%%%%%%%%%%%%%%%%%%%%%%%%%%%%%%%%%%%%%%%%%%%%%%%%%%%%%%%%%%%%%%
%%%%%%%%%%%%%%%%%%%%%%%%%%%%%%%%%%%%%%%%%%%%%%%%%%%%%%%%%%%%%%%%%%%%%%%%%%%%%%%%%%%%%%%%%%%%%%%%%%%%
%%%%%%%%%%%%%%%%%%%%%%%%%%%%%%%%%%%%%%%%%%%%%%%%%%%%%%%%%%%%%%%%%%%%%%%%%%%%%%%%%%%%%%%%%%%%%%%%%%%%
%%%%%%%%%%%%%%%%%%%%%%%%%%%%%%%%%%%%%%%%%%%%%%%%%%%%%%%%%%%%%%%%%%%%%%%%%%%%%%%%%%%%%%%%%%%%%%%%%%%%
%
%
%@@@@@@@@@@@@@@@@@@@@@@@@@@@@@@@@@@@@@@@@@@@@@@@@@@@@@@@@@@@@@@@@@@@@@@@@@@@@@@@@@@@@@@@@@@@@@@@@
%################################################################################################
%********************************** DERKACZ *********************************************
%\paragraph{Corncrake and Crickets:}
Finally, we checked amplitude fluctuations with time for the sound of a corncrake emitted from a loudspeaker toward the artificial meadow~--~fig.~\ref{fig:derkacz_OriginalVSn200}. The greatest changes in the amplitude (red), compared to the original one (blue), were observed for the moment when the shutter passed at the nearest distance from the emitting center. However, the main feature of the spectrum is that the signal is shifted in time and its base is elongating, i.e. the duration of each signal is prolonged; furthermore, its amplitude increases due to the interference effect.
\\
%%%%%%%%%%%%%%%%%%%%%%%%%%%%%%%%%%%%%%%%%%%%%%   SWIERSZCZ    %%%%%%%%%%%%%%%%%%%%%%%%%%%%%%%%%%%%%%%%%%%%%%%%%%%
Contrary to this, in the case of the cricket, the amplitude of the reflected sound rebuilds poorly compared to the level of the original signal and never reaches its maximum value. Nevertheless, similarly to the case of the corncrake, the amplitude of the signal fluctuates during the transition of the newcomer (fig.~\ref{fig:swierszcz_OriginalVSn200}). This spectrum reveals some more interesting findings. Some peaks here are not reconstructed and some irregular shifts in time are observed. The differences in the position of the peaks vary between the different phases of the experiment. We checked numerically, for example, that for a given density of blades, the shifts for the beginning and the middle part of the experiment differ by tenths of milliseconds. This interesting feature shows that the sound peaks are reconstructed by interference and are not a simple reflection of the original signal with the echo effect delay. Such an effect was observed many times and seems to be very common~--~see fig.~\ref{fig:05Hz1kHz} as an example (here, positions of minimum values remain the same, only maximum values change their positions).
\\
We were unable to reconstruct the parameters of a real meadow that is characteristic for a corncrake or a cricket in the studio. Thus, our artificial meadow made of dry cane would perhaps be inadequate to some extent for the sound of real animals that we used and the position of the peak of the sound is somewhat not reliable. However, the general finding of the dependence of the position of the peak on density is correct and should be explored in detail in the future.
%%%%%%%%%%%%%%%%%%%%%%%%%%%%%%%%%%%%%%%%%%%%%%%%%%%%%%%%%%%%%%%%%%%%%%%%%%%%%%%%%%%%%%%%%%%%%%%%%%%%%%
%%%%%%%%%%%%%%%%%%%%%%%%%%%%%%%%%%%%%%%%%%%%%%%%%%%%%%%%%%%%%%%%%%%%%%%%%%%%%%%%%%%%%%%%%%%%%%%%%%%%%%
%%%%%%%%%%%%%%%%%%%%%%%% NUMERICAL RESULTS %%%%%%%%%%%%%%%%%%%%%%%%%%%%%%%%%%%%%%%%%%%%%%%%
%%%%%%%%%%%%%%%%%%%%%%%%%%%%%%%%%%%%%%%%%%%%%%%%%%%%%%%%%%%%%%%%%%%%%%%%%%%%%%%%%%%%%%%%%%%%%%%%%%%%%%%%%%%%%%%%%%%%%%%%%%
\section{Numerical experiments}
\begin{figure}
\centering
\includegraphics[scale=0.45]{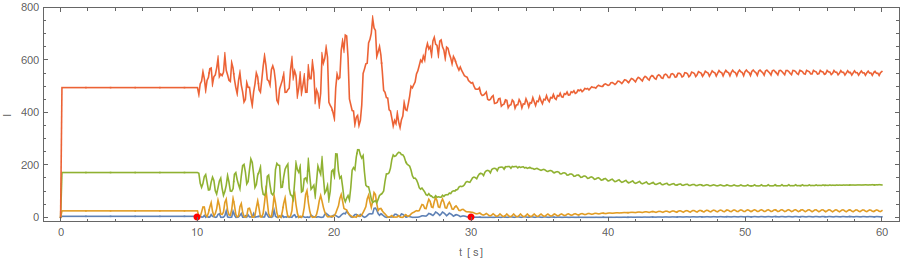}
\caption{Time-line of a numerically calculated relative intensity of a reflected 3.4~kHz tone emitted toward the meadow ($R$=1~m) for different values of the nearest neighbor distance $\alpha$, which are in meters: $\alpha$=0.3 (blue; almost no blades of the meadow), 0.13 (yellow), 0.07 (green), and 0.03 (orange). The velocity of the newcomer is 0.1~m/s, whereas the number of its compartments is 5 here. The red dots at the abscissa denote the time when the newcomer entered and left the meadow ring.}
\label{fig:InterferenceOfTone1}
\end{figure}

\begin{figure}
\centering
\includegraphics[scale=0.2]{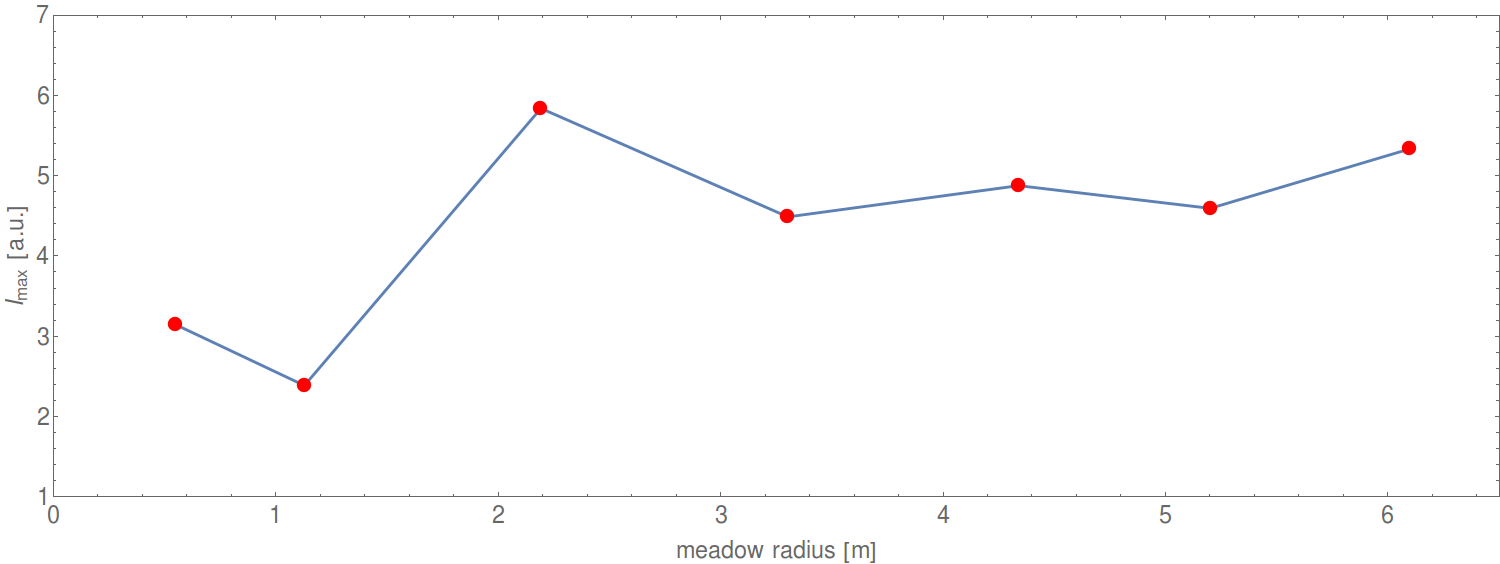}
\caption{Numerically calculated values of the relative intensity of the sound returning to the singer as a function of the radius of the meadow. $\alpha$ is kept at the value of 0.07~m.}
\label{fig:R_change}
\end{figure}

\begin{figure}
\centering
\begin{subfigure}{\linewidth}
\includegraphics[width=\textwidth]{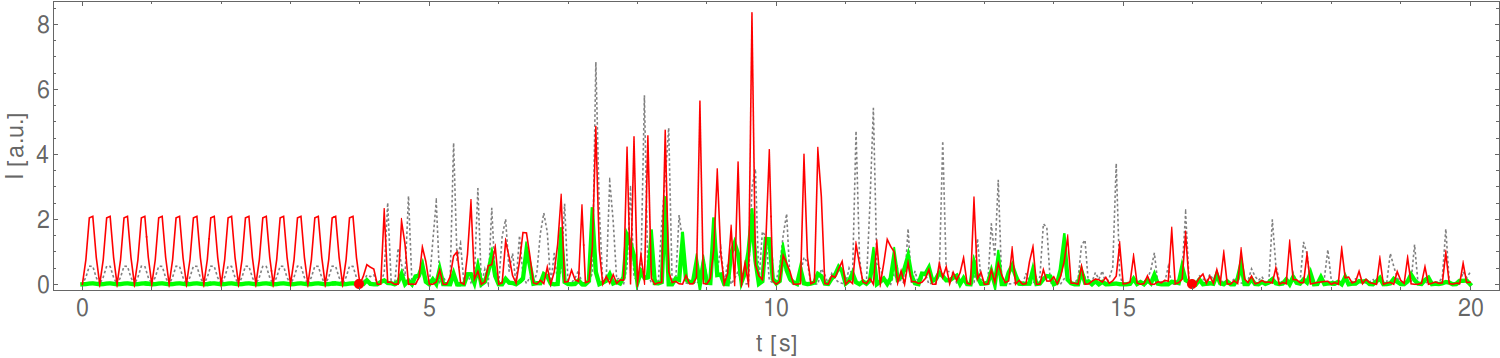}
\end{subfigure}
\begin{subfigure}{\linewidth}
\includegraphics[width=\textwidth]{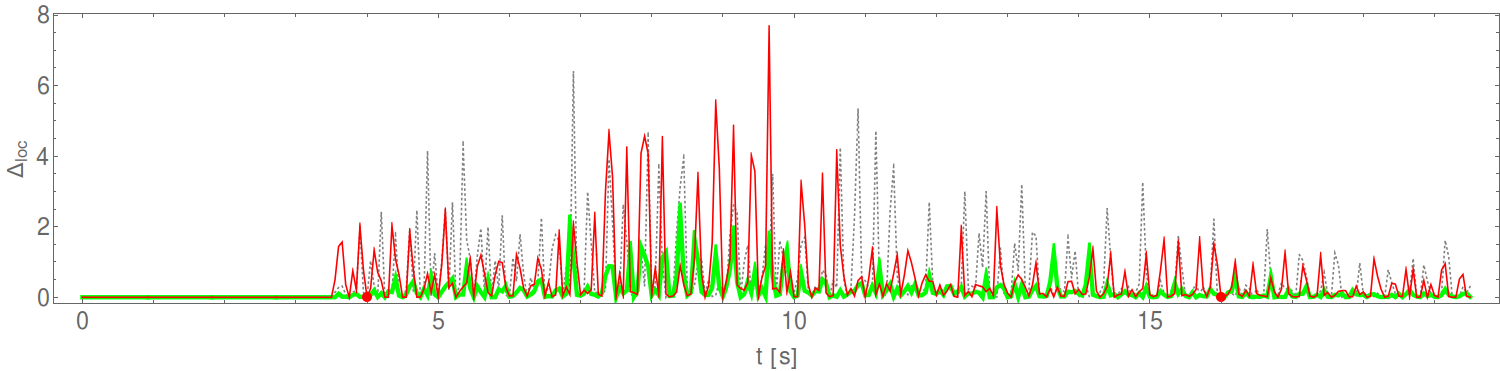}
\end{subfigure}
\begin{subfigure}{\linewidth}
\includegraphics[width=\textwidth]{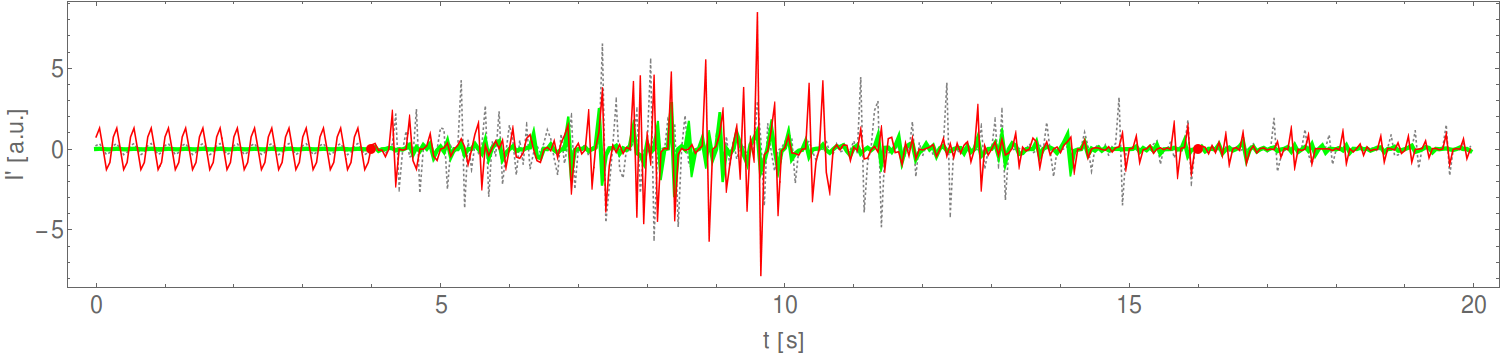}
\end{subfigure}
\caption{Numerically calculated values of the intensity of the reflected sound during passage of the newcomer across the meadow for several frequencies: $f$=1~kHz~--~gray dotted line, 5~kHz (optimal for the density of a meadow)~--~red line, 10~kHz~--~green thick line. The red points at the abscissa indicate the time of starting to move by the newcomer and the time of moving out of the meadow. In these numerical calculations, $R$=6~m, $f_m$=2~Hz, and $w$=1~m/s. The figure in the middle shows a local distortion function of the adequate signals, whereas the figure at the bottom is the 1$^{\text{st}}$ derivative of the intensity.}
\label{fig:f_change}
\end{figure}

\begin{figure}
\centering
\includegraphics[scale=0.3]{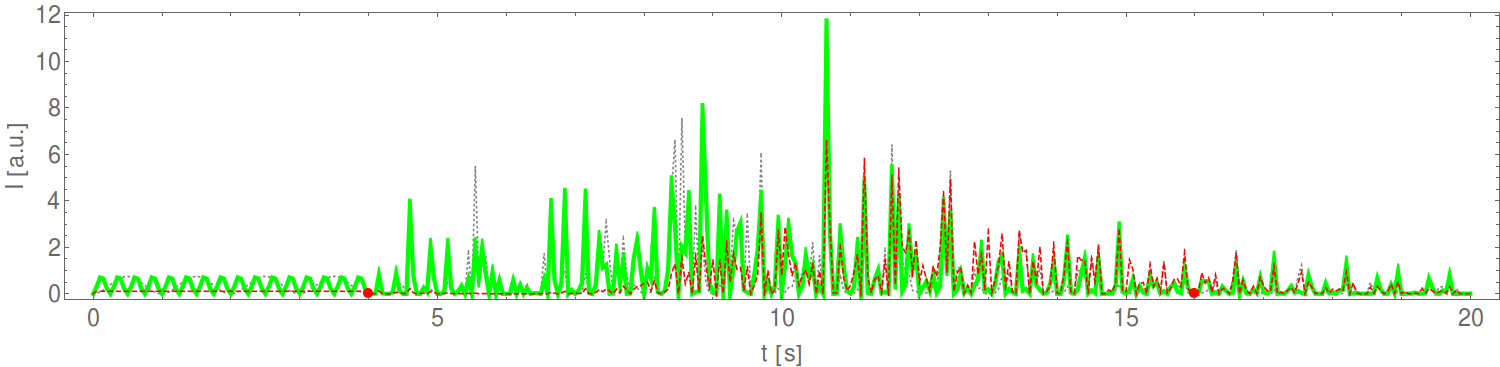}
\caption{Intensity of the reflected signal as a function of the modulation frequency $f_m$: 0.5~Hz -- gray dotted line, 2~Hz -- green line, 10~Hz -- red dashed line. The red points at the abscissa indicate the time of starting to move by the newcomer and the time of moving out of the meadow. Other parameters are $R$=6~m, $f$=5~kHz, and $w$=1~m/s.}
\label{fig:fm_change}
\end{figure}

\paragraph{Theoretical model.}
By similarity to the voice of crickets and a corncrake, we assumed that a representative of an animal voice can be a tone of frequency $f$ modulated by frequency $f_m$ (amplitude modulation). Numerically, the $f$ value is a representation of the pitch of the voice (e.g. 5~kHz), whereas $f_m$ represents the frequency of repetitions of the sound (chirps; $f_m$=2~Hz).\\
The wave emitted by the $i$-th singer is
\begin{multline}
\Psi_{i}(\boldsymbol{r},t)=\sin{\lbrack \Omega \cdot t-\boldsymbol{k} \circ (\boldsymbol{r}-\boldsymbol{r}^{(i)})\rbrack} \times \sin{\lbrack\Omega_m \cdot t-\boldsymbol{k}_m \circ (\boldsymbol{r}-\boldsymbol{r}^{(i)})\rbrack}= \\
=\sin{\Big( 2\pi f \cdot t- \frac{2\pi f}{c} \sum_{j=1}^3 (r_j-r_j^{(i)})\Big)}\ \times\ \sin{\Big( 2\pi f_m \cdot t-\frac{2\pi f_m}{c} \sum_{j=1}^3 (r_j-r_j^{(i)})\Big)},
\label{eq:FalaSpiewak}
\end{multline}
where $c$ is the speed of sound and $\boldsymbol{r}^{(i)}$ is the position vector of the $i$-th singer.\\
Furthermore, a wave incoming to a blade stimulates it to vibrate. The wave emitted by the $j$-th blade is a composition of spherical waves with frequency $f_1$ (as the simplest case, only one frequency was taken into account) and with an amplitude generated each time as a sum of $i$ amplitudes of incoming waves
\begin{equation}
\psi^{j}(\boldsymbol{r},t)=\sum_i^N \frac{\psi_0^{ij}(t)}{|\boldsymbol{r}-\boldsymbol{r}_j|} \sin{\Big(2\pi f_1\ (t-t_0^{ij})-\frac{2\pi f_1}{c} \sum_{k=1}^3{(r_k-r_k^{(j)})}\Big)} \cdot \theta(t-t_0^{ij}),
\label{eq:FalaZdzblo}
\end{equation}
where $\psi_0^{ij}(t)=\sqrt{\Psi_i^2(r_j,t)}$ is the wave amplitude at time $t$ generated by the $i$-th cricket in the place of the $j$-th blade, $N$-- is the number of crickets, $r^{(j)}$ is the position vector for the $j$-th blade, $\theta(t-t_0^{ij})$ is the step function, and $t_0^{ij}$ is the time at which the stimulating wave from the $i$-th cricket arrives at the $j$-th blade. The latest term is added because the $j$-th blade is stimulated to vibration by the $i$-th wave, when it arrives at it (but not before). The waves emitted from crickets influence not only the amplitude of vibration of a given blade. Also other blades which scatter spherical waves are important here. Therefore, in our code, the sum in eq.~(\ref{eq:FalaZdzblo}) does take into account other blades as a source of stimulating waves.\\
Waves defined by expression~(\ref{eq:FalaZdzblo}) propagate and reach back the $k$-th cricket. The wave at this place is described by
\begin{equation}
\psi_k(\boldsymbol{r}_k,t)=\sum_{j}\psi^{j}(\boldsymbol{r}_k,t).
\end{equation}
This provides the following expression for the sound intensity at the $k$-th cricket
\begin{equation}
I_k(t)=\sqrt{\psi^2(\boldsymbol{r}_k,t)}.
\label{eq:Intensity}
\end{equation}
%It is known that for many species the changes with time of a visual or acoustic signal play a more important role than a quality of this picture at given time. The Authors idea is to show that the greatest disturbance is for a signal adjusted to the parameters of a meadow. The adjusted sound is emitted by meadow animals who can utilise it as a tool which gives the information indicating changes.
%The signals which we are interested in are periodic ones in most cases. According to this, we introduced a quantitative measure of correlations called \emph{distortion}, $\Delta$. It is defined as a sum of differences between an intensity at given time and this for a time shifted by a period. A summation is taken over all signal samples which have a counterpart sample in a period further (only samples from a last period are lost since they have not a counterpart).
%%%%%%%%%%%%%%%%%%%%%%%%%%%%%%%%%%%%%%%%%%%%%%%%%%%%%%%%%
\paragraph{Description of the numerical experiments.}
Loudness as a function of time for a reflected sound from a meadow could be better understood by analyzing the numerical experiments that we conducted. Here, the meadow was simulated as a set of 2-dimensional points scattered through a circular area (fig.~\ref{fig:ArtificialMeadow}). Each of the blades of this artificial meadow is a center of circular waves emitted as a consequence of stimulation thereof by an incoming acoustic wave. This reduction of a problem to a 2-dimensional case is not primarily important to the main ideas presented here but does reduce time and resources consumed by numerical calculations considerably.\\
The newcomer was simulated here as a set of points (segments) which can reflect the sound waves as the blades can. The coordinates of the newcomer's segments were chosen randomly along the direction of its future movement. The distance between the segments was short, compared to the radius of the meadow (less than 1\% of $R$; fig.~\ref{fig:ArtificialMeadow}), and adequately to simulate an elongated and irregular shape of a worm.\\
We made programs allowing us to simulate the crucial acoustic features of a meadow. First, an artificial meadow was created. To do this, input parameters such as meadow's radius and the number of blades were set. Based on these parameters, the coordinates of blades and a newcomer were chosen randomly inside the radius by the software. The formula~(\ref{eq:Torquato}) was used to calculate the number of randomly scattered blades required to have given $\alpha$ for a meadow of a given radius $R$. As a result, we obtained a set of blades scattered quasi-uniformly within this circular field.
% Some of the numerical values calculated from this formula are given in tab.~\ref{tab:GestoscTorquato}.
%\begin{table}[h]
%\centering
%\begin{tabular}{|c|c|c|c|c|c|}
%\hline
%$N=$ & 100 & 200 & 300 & 400 & 500 \\
%\hline
%\begin{sideways}a [cm] \end{sideways} & 9 & 6 & 5 & 4 & 4 \\ \hline
%\end{tabular}
%\caption{Mean nearest neighbour distance $a$ for a meadow of radius 1~m with $N$ blades.}
%\label{tab:GestoscTorquato}
%\end{table}
%%In all simulations one dimension is skipped, i.e. there is no height -- here blades are reduced to points (but not rods) and thus a spherical wave is replaced by a circular one. This simplification seems to not lose the essence but makes simulations less time consuming.\\
In these numerical calculations, sound reflection coefficients for both the newcomer and the blades were set as input parameters. Finally, the blades' and newcomer's coordinates were used to calculate the intensity of sound as a function of time in the place of an emitting singer. To calculate this intensity, we used the formula~(\ref{eq:Intensity}). As a simplification, only one singer was considered (placed at point (0,0)~--~see fig.~\ref{fig:ArtificialMeadow}).
\paragraph{Results of the numerical experiments.}
The importance of the presence of blades and the interference effect is clear when the intensity for different densities of the meadow is compared. The blades reflect sound and thus they increase the intensity of the interfered sound returning to the singer. The left red dot on the abscissa in fig.~\ref{fig:InterferenceOfTone1} means a moment when the newcomer started to move, while the right one means the time when the newcomer left the area with blades and continued to move way. The data series in this figure show the results for different nearest neighbor distances $\alpha$. When the density of blades is much lesser than the wavelength (blue and yellow lines in this figure), the signal has a small amplitude. For $\alpha$ equal or less than the wavelength ($\alpha \le$ 0.07~m, in this case), the intensity of the reflected signal and its fluctuation amplitude increase considerably due to the interference.
\\
Two preliminary numerical experiments should be mentioned. First, we checked the variation of the loudness at the same $\alpha$ when the position of particular blades is randomly rearranged.
%It was made some trials of setting the random positions of blades but the density of blades and the parameters of the sound were kept.
Some trials of the positions of the blades were made for a meadow radii $R$=3~m and 8~m. In these experiments, we recorded a maximum of the intensity when the newcomer was absent. The conclusion is that, for both radii, the loudness of the back signal differs by 2-5~dB for different randomly chosen positions of blades.
%Symulacje Bartka 13VI2017
%/home/marek/Dokumenty/WhatILike/WoodsAndGrasses/Numerics/c/ObliczeniaZUczniami/SymulacjeBartka/VI_2017
\\
The next preliminary experiment consisted in recording of the sound for some radii $R$ of the meadow in a range of 0.5~m -- 6~m with a step of 0.5~m (keeping the positions of blades, density of the meadow, and parameters of the sound unchanged). It is expected that the influence of distant blades for a sufficiently wide meadow is negligible. In this case, the variation of loudness was 2-3~dB over all radii. It seems that the loudness of the sound for $R$=3~m is comparable to that for $R$=10~m~--~fig.~\ref{fig:R_change}. Thus, next experiments with changing other parameters of sound were performed for a meadow with $R$=6~m.
%Symulacje Bartka 7VI2017 i wczesniejsze
%/home/marek/Dokumenty/WhatILike/WoodsAndGrasses/Numerics/c/ObliczeniaZUczniami/SymulacjeBartka/VI_2017
\\
The main goal of our numerical simulations was to check how the periodic signal emitted by a cricket changes if some of the crucial parameters such as $f$, $f_m$, or $w$ vary. For each signal, we calculated and plotted some secondary functions of the intensity with time. They are chosen as intuitive functions represented changes in time.
%~--~fig.~\ref{fig:NumParams}
The most useful picture of loudness fluctuation is given by the first derivative of the intensity and the local distortion of the loudness. The distortion function $\Delta$ is a measure of changes of the intensity~--~for details see the Appendix. The larger is the distortion the greater are fluctuations of the intensity.\\
The $\alpha$ to $\lambda$ adjustment (it can be translated to the adjustment of the density of a meadow to the wave length) seems to be crucial when one considers a time function of the sound intensity.
%%{figs/AnalysisExampleForPaper_Part3_Derivative.png}
%\begin{figure}[ht] 
%  \begin{subfigure}[b]{0.5\linewidth}
%    \centering
%    \includegraphics[width=0.95\linewidth]{figs/AnalysisExampleForPaper_Part1_Intensity.png} 
%    \caption{Intensity} 
%    \label{fig7:a} 
%    \vspace{4ex}
%  \end{subfigure}%% 
%  \begin{subfigure}[b]{0.5\linewidth}
%    \centering
%    \includegraphics[width=0.95\linewidth]{figs/AnalysisExampleForPaper_Part2_Loudness.png} 
%    \caption{Loudness} 
%    \label{fig7:b} 
%    \vspace{4ex}
%  \end{subfigure} 
%  \begin{subfigure}[b]{0.5\linewidth}
%    \centering
%    \includegraphics[width=0.95\linewidth]{figs/AnalysisExampleForPaper_Part3_Derivative.png} 
%    \caption{Intensity derivative} 
%    \label{fig7:c} 
%  \end{subfigure}%%
%  \begin{subfigure}[b]{0.5\linewidth}
%    \centering
%    \includegraphics[width=0.95\linewidth]{figs/AnalysisExampleForPaper_Part4_Distortion.png} 
%    \caption{Distortion} 
%    \label{fig7:d} 
%  \end{subfigure} 
%  \caption{Some useful parameters of the signal}
%  \label{fig:NumParams} 
%\end{figure}
%Instead a density of a meadow $\sigma$ we will quote a mean nearest distance of grass blades, $\alpha$, as a representative of a meadow parameters. Both, $\sigma$ and $\alpha$ are related each other by the formula (\ref{eq:Torquato}).
For given $\alpha$, a given set of sound parameters may be regarded as well or more badly adjusted to signalling a move. We define a sound wave as a well-adjusted one to $\alpha$ if the wavelength $\lambda=c/f$ value of the sound is comparable to $\alpha$. Oppositely, when the wavelength and the $\alpha$ diverge considerably, the sound is more badly adjusted.\\
All the results below were obtained for $\alpha$=0.07~m (quite a realistic value). This distance equals to the wavelength of the tone $f$=$f_0\equiv$5~kHz characteristic for a corncrake or a cricket. The results were compared to frequencies $f<<f_0$ and $f>>f_0$. In these experiments, the newcomer is represented by 20 compartments. The sound reflection coefficient of the newcomer compartments was set equal to that for the blades of grass and set to 0.8 (its value does not change the results qualitatively). In each cases, the newcomer started to move 2~s after starting sound emission. The sampling step was 0.05~s.
For simplicity, the intensity rather than the loudness is presented\footnote{In the loudness definition $L \textrm{[dB]}=10\ log(I/I_0)$, the $I_0$ could be replaced for our purposes by the intensity of some referred tone instead of $I_0$ known for a standard definition of loudness related to human hearing capabilities, for example, the mean intensity of a sound reflected when a newcomer is at rest.}.\\
%%%%%%%%%%%%%%%%%%%%%%%%%%%%%%%%%%%%%%% f change %%%%%%%%%%%%%%%%%%%%%%%%%%%%%%%%%%%%%%%%%%%%%%%%%
Fig.~\ref{fig:f_change} presents changes in the sound from the meadow with time for modulated tones: $f$=1~kHz, 5~kHz, and 10~kHz, whereas $f_m$=2~Hz and $w$=1~m/s. Based on this example, a general regularity is shown~--~for a badly-adjusted frequency of $f>f_0$=5~kHz, the intensity of the signal decreases (the green line). That is the reason that the sound with a great frequency does not suffice to serve as a tool for indicating changes.
%/home/marek/Dokumenty/WhatILike/WoodsAndGrasses/Numerics/c/ObliczeniaZUczniami/SymulacjeBartka/V_2017/check_f_DeltaLocal.png
In contrast, the intensity and its fluctuations for a sound with $f<f_0$ seem to be more adequate from this point of view. However, for such a small frequency $f$ (gray line), the fluctuations are spread in time and are generally lesser than the fluctuations for $f=f_0$. The distortion function (see the Appendix) for these cases are $\Delta$=7.20, 7.24, and 6.84 for $f$=1, 5, and 10~kHz, respectively, which means that the greatest amplitude variation is for $f\sim f_0$.
\\
%%%%%%%%%%%%%%%%%%%%%%%%%%%%%%%%%%%%%%%%% f_m change %%%%%%%%%%%%%%%%%%%%%%%%%%%%%%%%%%%%%%%%%
The modulation frequency $f_m$ seems to play a non-trivial role in the intensity variation~--~fig.~\ref{fig:fm_change}.
Although the distortion varies slowly with $f_m$, its maximum value is at some intermediate $f_m$ ($f_m$=2~Hz here). Its value for $f_m$=0.5, 2, and 10~Hz are 7.27, 7.33, and 7.16, respectively. The maximum of the amplitude fluctuation as a function of $f_m$ is probably related to the velocity of the newcomer. This issue, however, has not been explored systematically yet.\\
Finally, we checked the influence of the velocity $w$ of the newcomer. It is clear that there are no interference changes in time if $w$=0. In general, the distortion is a weakly changing function of the velocity value in a wide range.
%~--~fig.~\ref{fig:w_change}
For example, for $f$=5~kHz, $f_m$=2~Hz, the $\Delta$ is 7.44, 7.45, and 7.45, for $w$=0.1~m/s, 0.5~m/s, 1~m/s, respectively.
%\begin{figure}
%\centering
%\includegraphics[scale=0.3]{figs/check_w_powt.png}
%\caption{The intensity of a modulated signal $f$=5~kHz, $f_m$=2~Hz with a velocity $w$ change is presented: 0.1~m/s -- grey dotted line, 0.5~m/s -- green line, 1~m/s -- red line.}
%\label{fig:w_change}
%\end{figure}
%
\\
The number of the compartments of the newcomer is relatively small compared to the number of blades (ca. 6$\times$10$^3$ here). Thus, the amplitudes of peaks are not large. By adding the compartments (or, to some extent, increasing the reflection coefficient), one can obtain much greater intensity fluctuations. The fluctuations of loudness that we observed in the acoustic experiments are easy to be simulated by increasing the size of the newcomer. For example, let us assume that ca. 400 blades form a meadow with $R$=2~m and $\alpha$=0.07~m. If the newcomer is made of $N_c$=10 compartments, the maximum intensity is 3$\times$10$^5$, in turn, if $N_c$=50, the maximum value is 8$\times$10$^7$.\\
To generalize, our numerical simulations prove the existence of amplitude (intensity) fluctuations during a passage of the object reflecting sound throughout the meadow. The greatest fluctuations are obtained for a sound with the frequency well-adjusted to $\alpha$.
%%%%%%%%%%%%%%%%%%%%%%%%%%%%%%%%%%%%%%%%%%%%%%%%%%%%%%%%%%%%%%%%%%%%%%%%%%%%%%%%%%%%%%%%%%%%%%%%%%%%%%%%%%%%%%%%%%%%%%%%%%
%%%%%%%%%%%%%%%%%%%%%%%%%%%%%%%%%%%%%%%%%%%%%%%%%%%%%%%%%%%%%%%%%%%%%%%%%%%%%%%%%%%%%%%%%%%%%%%%%%%%%%%%%%%%%%%%%%%%%%%%%%
%\section{Reactions of bred crickets on motion around}
%$\leadsto$ (\textsc{Grzesiek \& Me}) We bred crickets ...\\
%... and observed that they stopped (???) to chirp when we were making a move of a shutter in the darkness.
%%%%%%%%%%%%%%%%%%%%%%%%%%%%%%%%%%%%%%%%%%%%%%%%%%%%%%%%%%%%%%%%%%%%%%%%%%%%%%%%%%%%%%%%%%%%%%%%%%%%%%%%%%%%%%%%%%%%%%%%%%
\section{Conclusions}
We have shown both experimentally and by numerical experiments that the sound emitted by an animal living in grass or shrubs returns to this animal after reflections from blades of the meadow and interference and forms a complex pattern of changes in the intensity with time. This picture of changes is sensitive to some crucial parameters of the sound and the meadow like the velocity of the newcomer or the relation between the sound frequency and the density of the meadow. The relative variation in the loudness is some decibels. Such a variation of the intensity should be distinguishable and allows regarding these fluctuations as a possible tool that could be used by animals for detection of movement.\\
Hearing distortions seem to be stable within the range of natural changes in temperature~--~when the temperature changes by ten degrees or more, the sound wavelength changes by millimeters, and the interference conditions remain almost unchanged. Also, the changes of the rate of chirp with temperature (Dolbear's law~\cite{Dolbear}) does not affect the quality of this tool (fig.~\ref{fig:fm_change}).\\
Sensing the environment by analyzing sound amplitude changes seems to be an interesting issue (according to our knowledge, not reported until now) from both the basic-knowledge and ecological points of view. For example, if the mechanism of such detection of a mate or a predator is really used by living creatures, some modification of the environment by human activity could change the size of a population. However, one should be aware that sound communication in real biological systems depends on biocoenotic and geological conditions. Both of them should be incorporated as parameters in a mathematical model of the acoustic environment of this system. Here, the basic version of such a model is presented.\\
In this paper, we limited the investigations to the case when only one sound-emitting animal is taken into account. Nonetheless, it would be interesting to consider a net of animals which cooperate with each other in producing a field of sound for detection the motion in the surroundings.
\section{Acknowledgement}
The authors want to thank A.~Nieoczym, the manager of 'Chatka \.Zaka' and M.~Skowronek, the manager of \emph{Radio Centrum} for enabling us to perform the acoustic experiments in a recording studio.\\
Thanks for Prof.~B.~Bernatowicz (Department of Art of UMCS University) for enabling us to perform our first experiments in an acoustically controlled space.\\
We (B.T. and J.M., especially) would like to thank Mr. P.~Kononowicz (Olympic special-interest group for young physicists in Lublin) for developing passion for discovering Nature's marvels in young people.
\section{Appendix~--~Distortion function}
\begin{figure}
\centering
\includegraphics[scale=0.8]{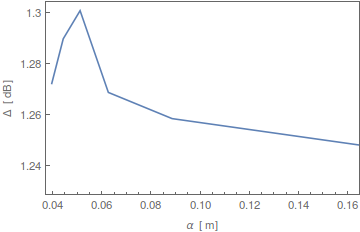}
\caption{Distortion function $\Delta$ of the prepared sound of a corncrake as a function of the nearest neighbor distance $\alpha$.}
\label{fig:DistortionCorncrakeExperimental}
\end{figure}

Because of the overlap of the position of many samples in the figures used in this paper (CD-quality sampling rate), real differences cannot be presented graphically. Therefore, we introduced a numerical measure for a periodic signal to quantify distortions made by a disturbing factor (like a shutter entering the meadow).
We define \emph{distortion} $\Delta$ as a sum of absolute differences in loudness $L_i$ for the $i$-th and $(i+T)$-th samples, which are distanced by one $T$ period, i.e. $\Delta=\sum_{i=1} |L_{i+T}-L_i|$. The summation is taken over all samples that have their counterpart at a period forward in the recording. Finally, to normalise the expression, this sum is divided by the number of samples\footnote{In this text, we also used a notion of the \emph{local distortion}, which is a distortion function in time by means of the time series of the absolute difference in the intensity at given time and that for a period forward.}. For example, for a periodic signal without disturbances, $\Delta$=0. The distortion from periodicity contains information about the presence of a disturbing factor.
\\
For instance, in the case of modulated ($f_m$=5~Hz) tones: $f$=0.5~kHz, 5~kHz, and 10~kHz, the value of $\Delta$ is 0.15, 4.1, 0.6 for $\alpha$=0.06~m and 0.17, 5.4, and 0.48 for $\alpha$=0.04~m, respectively. This means that the highest distortion for quite realistic densities of a meadow is noted in the case of a sound with an intermediate frequency.\\
% DISTORTION FOR ANIMALS
Similarly, the sound amplitude fluctuations for our representative animals were quantified by us using the $\Delta$. In these experiments, the starting and stopping time of the newcomer move was normalized (to tenths of a second) and the prepared sounds samples were used. For example, the $\Delta$ for the sound of a corncrake as a function of $\alpha$ is shown in fig.~\ref{fig:DistortionCorncrakeExperimental}.
%/home/marek/Dokumenty/WhatILike/WoodsAndGrasses/Numerics/c/ObliczeniaZUczniami/Dzwieki/RCentrum_11_10_2016/Jasiek_MiaraZnieksztalcenia/Plots
The peak of this experimental function shows that meadows with moderate density (the peak is at ca.~$\alpha$=0.05~m) are preferable to serve as a medium providing information about changes therein.
\bibliographystyle{ieeetr}
\bibliography{references}
\end{document}